\documentclass[aps,prb,twocolumn,superscriptaddress]{revtex4-1}
\usepackage{amsmath}
\usepackage{graphicx}
\usepackage{textcase}
\usepackage{bm}
\usepackage{epsfig}
\usepackage{natbib}

\begin{document}

\title{Numerical Simulation of the Nernst Effect in Extreme Type-II Superconductors:\\ A Negative Nernst Signal and its Noise Power Spectra}

\author{Sangwoo S.~Chung}
\affiliation{Physics \& Astronomy Department, Rice University, 6100 Main Street, Houston, TX 77005}
\affiliation{Department of Physics, University of Cincinnati, 345 Clifton Court, Cincinnati, OH 45221}

\author{Paata Kakashvili}
\altaffiliation[Present Address: ]{Department of Physics and Astronomy, Rutgers University, 136 Frelinghuysen Road, Piscataway, NJ 08854-8019}
\affiliation{NORDITA, Roslagstullsbacken 23, 106 91 Stockholm, Sweden}
\affiliation{Physics \& Astronomy Department, Rice University, 6100 Main Street, Houston, TX 77005}

\author{C.~J.~Bolech}
\affiliation{Department of Physics, University of Cincinnati, 345 Clifton Court, Cincinnati, OH 45221}
\affiliation{Physics \& Astronomy Department, Rice University, 6100 Main Street, Houston, TX 77005}

\date{\today}

\begin{abstract}

Recently, different transport coefficients have been measured in high-$T_{c}$ superconductors to pinpoint the nature of the pseudogap phase. In particular, the thermoelectric coefficients received considerable attention both theoretically and experimentally. We numerically simulate the Nernst effect in extreme type-II superconductors using the time-dependent Ginzburg-Landau equations. We report the sign reversal of the thermoelectric coefficient $\alpha_{xy}$ at temperatures close to the mean-field transition temperature $T^{\mathrm{MF}}_{c}(H)$, which qualitatively agrees with recent experiments on high-$T_{c}$ materials. We also discuss the noise power spectrum of $\alpha_{xy}$, which shows $1/f^{\beta}$ behavior. Based on this observation, we propose an experiment to distinguish among different regimes of vortex dynamics by measuring the noise correlations of the Nernst signal.
\end{abstract}

\maketitle

\section{Introduction}
The unexpected observation of unusually large Nernst coefficients ($\nu$) in cuprates at temperatures much greater than $T_c$ \cite{xu2000vortex} have drawn much spotlight to the Nernst effect over the past decade.  As the Nernst coefficients are generally small in most ordinary metals, \footnote{One notable exception being bismuth; see K.~Behnia, M.-A.~M\'easson, and Y.~Kopelevich, Phys.~Rev.~Lett.~{\bf 98}, 166602 (2007).} the large values of $\nu$ seen at $T > T_c$ have elevated the Nernst effect as one of the central avenues used in attempts to characterize the nature of the pseudogap state.  Since then, an extensive investigation on the subject has been done,  both experimentally \cite{xu2000vortex,wang2001,wang2006nernst,wang2002high,pourret2006observation,cyr2009enhancement,daou2010broken,hess2010,albenque,*albenque2} and theoretically,  \cite{ussishkin2002,mukerjee2004,podolsky2007nernst,raghu2008,tinh2009,andersson2010,alexandrov_nernst_2004,*tewari_effects_2009,*hackl_nernst-effect_2009,*hackl_quasiparticle_2010,kotetes2010} producing widely differing proposals on the origin of the phenomenon.  Most of these competing interpretations center in the dynamics of either vortices \cite{xu2000vortex,wang2001,wang2006nernst,wang2002high,ussishkin2002,mukerjee2004,podolsky2007nernst,raghu2008,tinh2009,andersson2010} or quasiparticles. \cite{cyr2009enhancement,daou2010broken,hess2010,alexandrov_nernst_2004,*tewari_effects_2009,*hackl_nernst-effect_2009,*hackl_quasiparticle_2010}

In this article, we use the time-dependent Ginzburg-Landau (TDGL) equations to simulate the Nernst effect for an extreme type-II superconductor and (i) report a sign reversal of the transverse thermoelectric coefficient $\alpha_{xy}$ consistent with the one observed in experiments, and (ii) examine the noise properties of the Nernst signal, which has been surprisingly overlooked thus far for the Nernst effect and, we find, can distinguish among different dynamical phases of the vortex system.

\begin{figure}[!htb]
\includegraphics[width=2.5in]{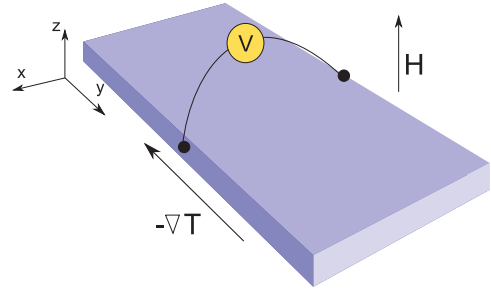} 
\caption{(Color online)  Geometry of a Nernst experiment.  A transverse voltage $V$ in the $x$ direction appears when a magnetic field $\mathbf{H}$ and a thermal gradient $\nabla T$ are applied simultaneously in the $z$ and $y$ directions, respectively.}
\label{nernstgeometry} 
\end{figure}

The Nernst effect is the emergence of a transverse voltage $V$ under the presence of an applied thermal gradient $\nabla T || \mathbf{\hat{y}}$ and a perpendicular magnetic field $\mathbf{H} || \mathbf{\hat{z}}$ (see Fig.~\ref{nernstgeometry}). From linear-response theory, the transport electric current due to an applied electric field $\mathbf{E}$ and a thermal gradient $\nabla T$ has the form $\mathbf{J} = \hat{\sigma}\mathbf{E} - \hat{\alpha}\nabla T$, where $\hat{\sigma}$ and $\hat{\alpha}$ are the electric and thermoelectric conductivity tensors.  Restricted to systems with negligible Hall effect, the off-diagonal term of $\hat{\alpha}$ is related to the Nernst coefficient as $\nu(H,T) \approx \alpha_{xy}(H,T)/H\sigma_{xx}(H,T)$.\footnote{Although $|\alpha_{xy}| \ne |\nu|$, $\mathrm{sgn}(\alpha_{xy}) = \mathrm{sgn}(\nu)$ with $H>0$.}
 
\section{Extreme type-II limit}
The TDGL equations can formally be derived by taking the Ginzburg-Landau free-energy functional \cite{tinkham} and employing Langevin dynamics for the time evolution of the superconducting order parameter $\psi$.  For quintessential high-$T_c$ superconductors, it is appropriate to consider the extreme type-II limit where $\kappa$, the ratio of the penetration depth $\lambda$ and the coherence length $\xi$, is assumed to be infinite by taking $\lambda \rightarrow \infty$.  In this simplified picture, the applied magnetic field in the sample is uniform and unscreened, as the lower critical field $H_{c1} \sim (\ln \kappa)/\kappa \rightarrow 0$ \cite{tinkham}.   Therefore, $\mathbf{H}(\mathbf{r},t)=H_{\mathrm{ext}}(t)\hat{\mathbf{z}}$ and we select an instantaneous Landau gauge for the vector potential $\mathbf{A}(\mathbf{r},t)=H_{\mathrm{ext}}(t)\left(\mathbf{r}\times\hat{\mathbf{z}}\right)/2$.  Consequently, the $\nabla \times \mathbf{H}$ term in Maxwell's equations vanishes, and we obtain the following relations that govern the evolution of $\psi(\mathbf{r},t)$ and the electric potential $\phi(\mathbf{r},t)$,
\begin{eqnarray}
\left(\partial_t+i\phi\right)\psi &=& -\frac{1}{\eta}\big[(-i\nabla-\mathbf{A})^2\psi \nonumber\\
&&+(1-T)\left(|\psi|^2-1\right)\psi\big]+\tilde{f}, \label{tdgl1}\\
\partial_t\mathbf{A}+\nabla\phi &=& (1-T)\mathrm{Re}\left[\psi^*(-i\nabla-\mathbf{A})\psi\right],\label{tdgl2}
\end{eqnarray}
along with zero perpendicular supercurrent boundary conditions, $(-i\nabla-\mathbf{A})\psi|_{\hat{\mathbf{n}}}=0$.

Temperature, length and time are measured in units of the mean-field transition temperature $T_c^{\mathrm{MF}}$, zero-temperature coherence length $\xi(0)$, and $t_0=\pi\hbar/(96k_BT_c^{\mathrm{MF}})$, respectively. Vector potential $\mathbf{A}$ and magnetic field $\mathbf{H}$ are in units of $H_{c2}(0)\xi(0)$ and $H_{c2}(0)$, respectively, where $H_{c2}(0)$ is the upper-critical field at $T=0$.  The dimensionless parameter $\eta$ is proportional to the ratio of characteristic times for $\psi$ and $\mathbf{A}$, and $\tilde{f}$ is a random thermal noise.  The order parameter $\psi$ has been normalized by the equilibrium value $\psi_{\infty}(T)$.

\section{Numerical Scheme}
\begin{figure}[!htb]
\includegraphics[width=2.5in]{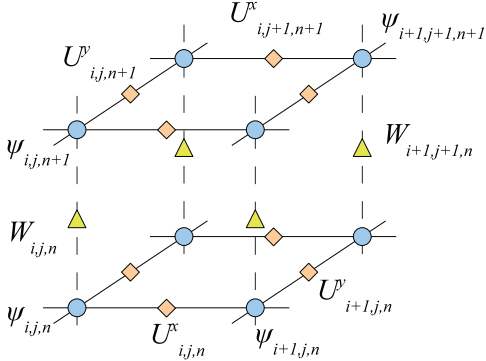} 
\caption{(Color online)  Scheme of discretization.  The order parameter $\psi$ is a nodal variable
and $U^x$, $U^y$, and $W$ are placed on links connecting the nodes.  We take the grid spacing $\Delta x=\Delta y=a=0.5$, while we set $\Delta t=0.015$ for the time interval between successive time steps,  both in normalized units.  These values were chosen so as to be able to resolve the structure and dynamics of the vortices.  Use of smaller values does not alter the results in any significant way.}
\label{Scheme} 
\end{figure}

Initially inspired by the Euler-forward based finite-difference scheme in Refs.~\onlinecite{kato,*kato2} and \onlinecite{bolech1995,*buscaglia2000} (cf.~Ref.~\onlinecite{Rosenstein}), we develop a gauge-invariant discrete implementation of Eqs.~(\ref{tdgl1}) and (\ref{tdgl2}).  Space and time have been discretized as shown in Fig.~\ref{Scheme} with $\psi_{i,j,n}$'s placed on the nodes, which are at fixed local temperatures $T_{i,j}$.  In addition to the spacelike link variables, $U^{x}_{i,j,n}=\exp(-i\int_{x_i}^{x_{i+1}}  d\xi A_{x}(\xi,y_j,t_n))$ and $U^{y}_{i,j,n}=\exp(-i\int_{y_j}^{y_{j+1}}  d\xi A_{y}(x_i,\xi,t_n))$, we introduce a time-like link variable $W_{i,j,n} = \exp\left(i\int_{t_n}^{t_{n+1}}d\tau \phi(x_i,y_j,\tau)\right)$ to incorporate $\phi$ into the numerics.

To write Eqs.~(\ref{tdgl1}) and (\ref{tdgl2}) in discrete forms, accurate to order $O(\Delta t^2)$ and $O(a^2)$, we first note that
\begin{eqnarray}
&U^{x}_{i,j,n}&W_{i+1,j,n}U^{x*}_{i,j,n+1}W_{i,j,n}^*\nonumber\\
&=& \exp\left(i\int_{x_{i}}^{x_{i+1}}d\xi\int_{t{}_{n}}^{t_{n+1}}d\tau
\left[\partial_{\tau}A_{\xi}+\partial_{\xi}\phi\right]_{y=y_j}\right)\nonumber\\
&\approx&1+ia\Delta t\left[\partial_tA_x+\partial_x\phi\right]_{x_{i+1/2},y_j,t_{n}},\label{uwuw}
\end{eqnarray}
where $x_{i+1/2}\equiv(x_{i}+x_{i+1})/2$.  Using Eq.~(\ref{tdgl2}) and the relation
$\mathrm{Re}\left[\psi^*(-i\partial_x-A_x)\psi\right]_{x_{i+1/2},y_j,t_n}\approx
\frac{1}{a}\mathrm{Im}\left(\psi^*_{i,j}U_{i,j}^x\psi_{i+1,j}\right)_n$, Eq.~(\ref{uwuw}) yields
\begin{eqnarray}
W_{i+1,j,n}&&=U^{x*}_{i,j,n}U^{x}_{i,j,n+1}W_{i,j,n}\times\nonumber\\
&&\Big[1+i\Delta t(1-T_{i,j})\mathrm{Im}\left(\psi^*_{i,j}U_{i,j}^x\psi_{i+1,j}\right)_n\Big],\label{wn}
\end{eqnarray}
and together with an analogous expression in the $y$ direction, they provide a solution to a Neumann-type boundary-value problem for $W_{i,j,n}$'s [with $\psi_{i,j,n}$'s given, and $U_{i,j,n}$'s fixed by $H_{\text{ext}}(t_n)$].

Moreover, it can be shown that Eq.~(\ref{tdgl1}) in terms of the discrete variables can be expressed as $\left(W_{i,j,n}\psi_{i,j,n+1}-\psi_{i,j,n}\right)/\Delta t=\mathcal{P}_{i,j,n}+\tilde{f}_{i,j,n}$ with
\begin{eqnarray}
\mathcal{P}_{i,j,n}=(1/\eta)\big[&&\big(U_{i,j}^{x}\psi_{i+1,j}-2\psi_{i,j}+U_{i-1,j}^{x*}\psi_{i-1,j}\big)/a^2\nonumber\\
&&+\big(U_{i,j}^{y}\psi_{i,j+1}-2\psi_{i,j}+U_{i,j-1}^{y*}\psi_{i,j-1}\big)/ a^2\nonumber\\
&&-\left(1-T_{i,j}\right)\left(\psi_{i,j}\psi_{i,j}^*-1\right)\psi_{i,j}\big]_{n},\label{pijn}
\end{eqnarray}
from which $\psi_{i,j}$'s at $t=t_{n+1}$ may be evaluated.

We perform our simulations on square grids with $N\times N$ cells with $N$ ranging from 64 to 384.  A time-independent thermal gradient is realized by linearly varying the local temperature $T_{i,j}$ along the $y$ direction of the sample.  For $N=192$, we apply a fixed temperature difference $\Delta T=0.03$ throughout, so that temperatures at the coldest and the hottest edges are $T_\mp=T\mp\Delta T/2$, respectively, where $T$ marks the average temperature of the sample.  The thermal fluctuation term $\tilde{f}$ is a Gaussian noise with standard deviation $\sigma_{i,j} = \sqrt{\pi E_o T_{i,j}\Delta t/6}$, where $E_o$ is a material-dependent parameter.  For our choice of $\eta = 1$, we set $E_0=1$ (guided by the fluctuation-dissipation theorem and values used in the literature to fit experiments \cite{mukerjee2004}) and confirmed that the results do not depend sensitively on variations around this value.

\section{Simulation results}
Our initial setting at $t=0$ is the Meissner state, where $\psi_{i,j}=U_{i,j}^x=U_{i,j}^y=1$.  With the fixed thermal gradient applied, we ramp up the magnetic field to a set value in the first few thousand steps and wait for 0.2 million time steps to reach a steady state.  We then time average $\alpha_{xy}(t_n)$ over 0.8 million time steps to obtain $\alpha_{xy}$, where
\begin{eqnarray}
\alpha_{xy}(t_n)=-\frac{\overline{J_x(t_n)}}{|\nabla T|}=-\frac{\iint J_x(x,y,t_n)dxdy}{Na\Delta T},\label{Jxavg}
\end{eqnarray}
and $J_x=(1-T)\mathrm{Re}\left[\psi^*(-i\partial_x-A_x)\psi\right]$ at $t=t_n$. \vspace{3mm}

\subsection{Transverse thermoelectric response}
\begin{figure}[!htb]
\includegraphics[width=3.3in]{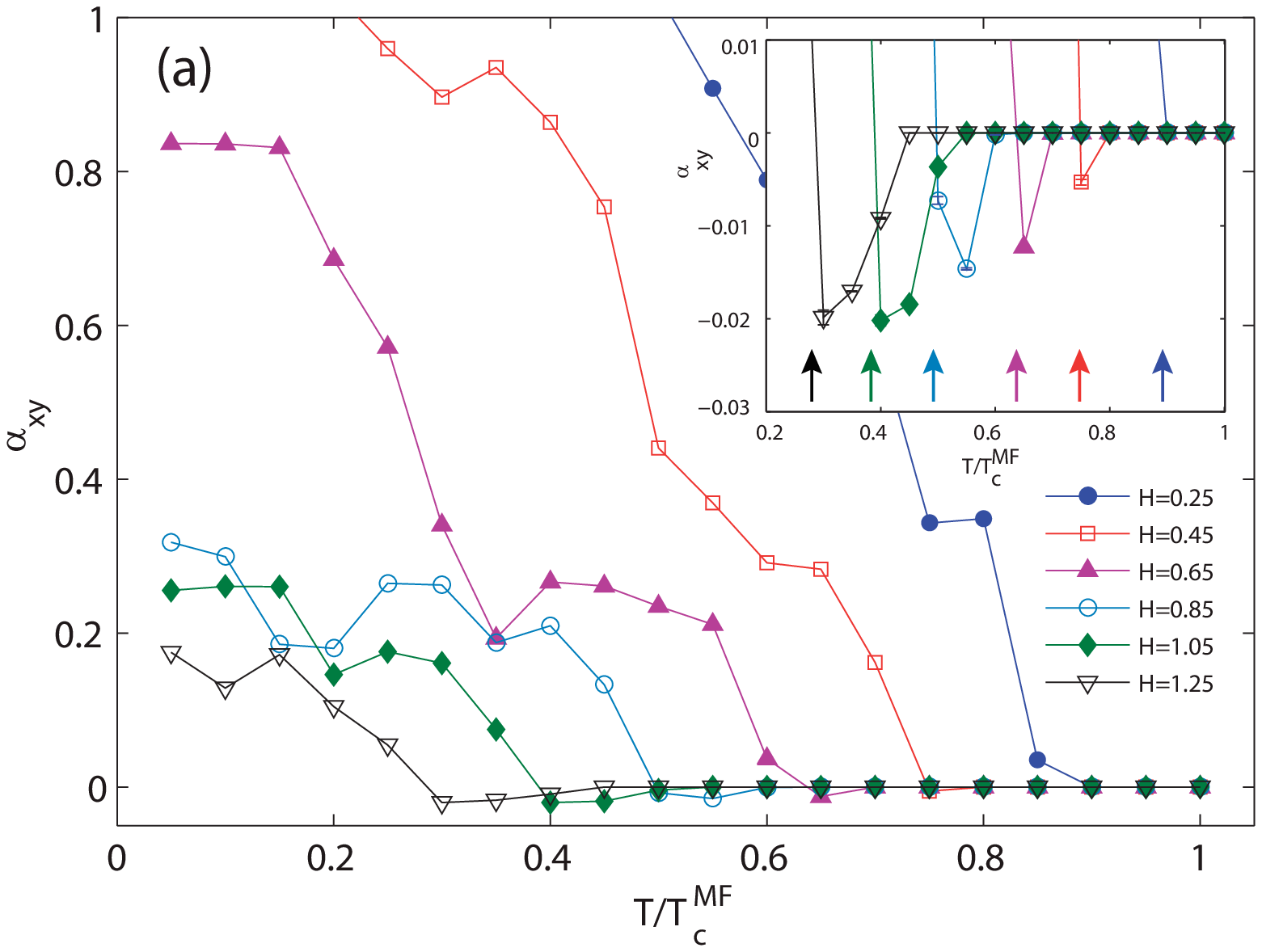}
\includegraphics[width=3.33in]{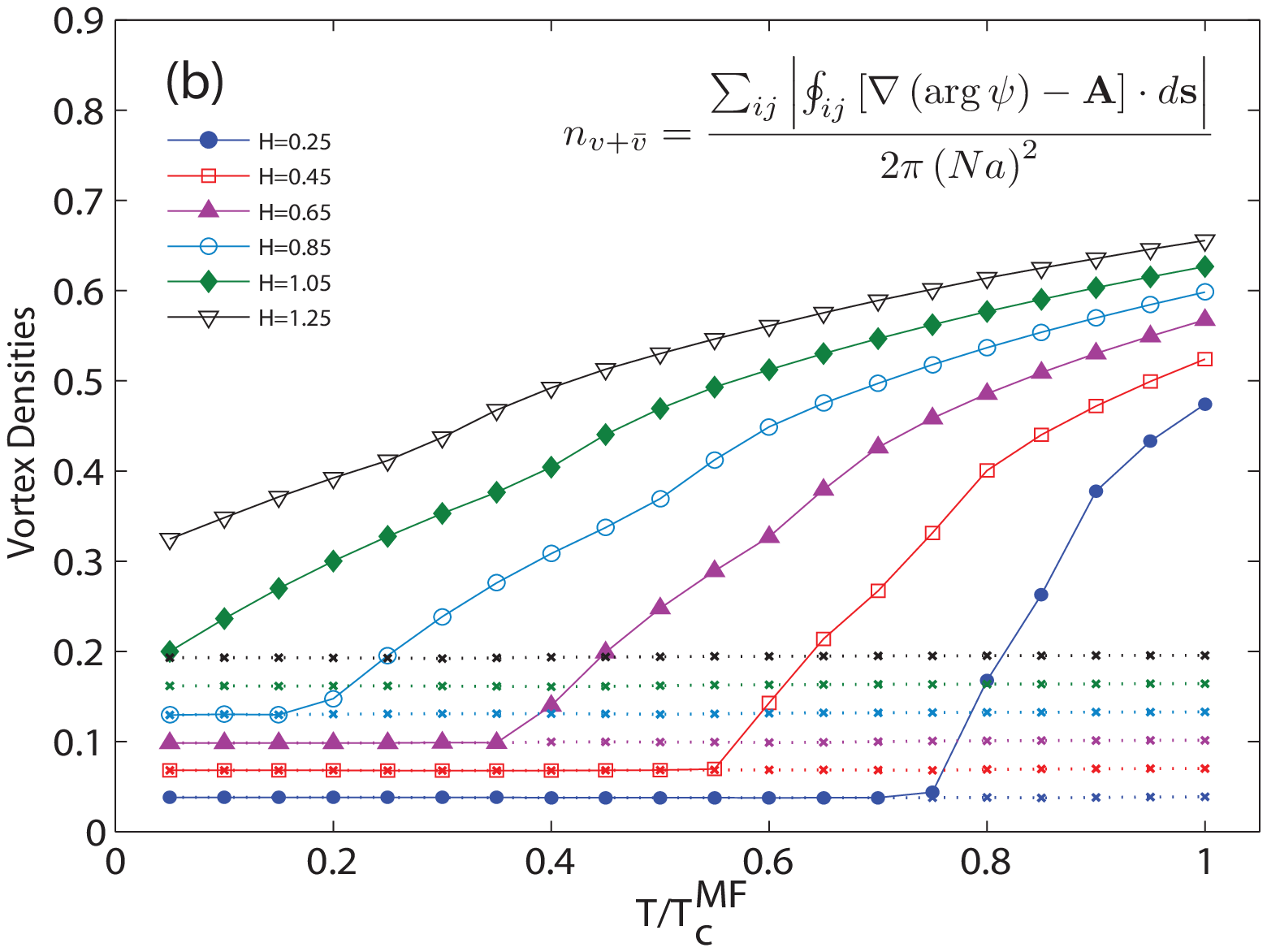}
\caption{(Color online)  (a) Temperature dependence of $\alpha_{xy}$.  Inset: $\alpha_{xy}$ zoomed into the negative-Nernst-signal region with $T_{\mathrm{NN}}(H)$'s marked by vertical arrows; the error bars (comparable to symbol sizes) highlight the statistical accuracy of the negative signals.  The error bar for each data point has been obtained from eight independent measurements of $\alpha_{xy}$, each measurement consisting of time averaging over 0.1 million time steps.
(b) Temperature dependence of $n_{v+\bar{v}}$, the total density of vortices plus antivortices, showing the onset of the vortex-proliferation regime, which is correlated to small kinks in $\alpha_{xy}$ and is seen to be unrelated to the onset of the negative Nernst signal.  The net vorticity $n_{v-\bar{v}}$ (dotted lines) reflects the applied magnetic flux that is independent of temperature.}
\label{Alphaxy}
\label{Vortices}
\end{figure}

Figure \ref{Alphaxy}(a) is the plot of $\alpha_{xy}$ vs temperature for various magnetic fields (see Fig.~\ref{alpha_xy_vs_H_negative} for $\alpha_{xy}$ vs $H$).  At low temperatures and, particularly, at lower magnetic fields, $\alpha_{xy}$ attains large positive values as also seen in both experimental \cite{wang2002high} and theoretical \cite{mukerjee2004, podolsky2007nernst} results.  As $T$ is increased,  $\alpha_{xy}$ falls monotonically until it reaches a vortex-proliferation temperature $T_{\mathrm{VP}}(H)$, which can be identified from Fig.~\ref{Alphaxy}(b)  (see Fig.~\ref{VorticityVsH} for vortex densities vs $H$).  At $T_{\mathrm{VP}}(H)$, the number of vortices and antivortices proliferate and the long-range phase coherence of $\psi$ is destroyed.  The exhibition of high $\alpha_{xy}$ values at $T<T_{\mathrm{VP}}$ is consistent with the very low resistances of the samples in that regime (cf.~Ref.~\onlinecite{emery1995importance}).  Above $T_{\mathrm{VP}}(H)$, the curves of $\alpha_{xy}$ resemble the profiles seen in Nernst coefficients $\nu$ for typical high-$T_c$ superconductors.  At even higher temperatures, $\alpha_{xy}$ becomes negative, and we designate the onset temperature for negative Nernst signals as $T_{\mathrm{NN}}(H)$ for later discussions.

Above $T_{\mathrm{NN}}$, we find that the onset of negative Nernst signal is correlated to the onset of the superconducting to normal transition at the hotter edge.  Prior to the transition, $H$ lies in the range $H_{c2}(T_-)<H<H_{c3}(T_+)$, where $H_{c3}$ is the critical field that destroys surface superconductivity.  The bulk is already at the normal state, while the Nernst signal is dominated by superconducting edge currents circulating around the bulk of the sample.  As $T$ is continually increased, $H_{c3}(T)$ decreases and the hotter edge crosses the phase boundary to enter the normal state.  Consequently, the Cooper-pair density at that edge is abruptly suppressed, resulting in a significant asymmetry in the supercurrent density relative to the colder edge and a different vortex vs antivortex dynamics across the sample.

From experiments, the negative $\alpha_{xy}$ reported in Ref.~\onlinecite{albenque,*albenque2} for an optimally doped YBCO sample displays a notable resemblance to our results.  A qualitatively similar profile of negative Nernst \emph{signals} can also be inferred from measurements of Nernst \emph{coefficients} done on a wide range of samples at various doping levels, \cite{wang2006nernst,daou2010broken} which indicates that this effect is ubiquitous and independent of details or fine tuning.  Theoretical results on granular superconductors in Ref.~\onlinecite{andersson2010} also show negative Nernst signals, which they interpret as diffusion of vortex vacancies in a pinned-vortex state.  In our simulations, however, the sign reversal is not from quasiparticle effects,\cite{alexandrov_nernst_2004,*tewari_effects_2009,*hackl_nernst-effect_2009,*hackl_quasiparticle_2010} as that scenario is not comprised in the Ginzburg-Landau framework.  Moreover, vortex vacancies do not seem to be the cause, since it can be inferred from Ref.~\onlinecite{andersson2010} that $T_{\mathrm{NN}}(H)$ would increase as $H$ is increased, while in Fig.~$\ref{Alphaxy}$(a) we see that $T_{\mathrm{NN}}(H)$ displays an opposite trend as the field is increased. \vspace{3mm}

\subsection{Noise analysis}
The dynamic phase diagram of vortex matter is more diverse than the one at equilibrium. \cite{koshelev_dynamic_1994,*giamarchi_moving_1996} In our simulations, at finite $\kappa$ and in the absence of pinning centers, we observe rich dynamical behavior.  Depending on the strength of the driving force (temperature gradient in our case), temperature, and magnetic field, vortex motion can be characterized into different regimes: immobile lattice of vortices due to the geometrical pinning in the finite-size sample, sliding vortex crystal, and flowing vortex liquids.  It is not surprising that noise properties of transport signals depend on the dynamical behavior of vortices.  Especially, since the Nernst signal is highly sensitive to vortex dynamics, we propose the measurement of its noise properties as a means to tell apart different dynamical scenarios. For this purpose, we have computed the noise correlation for $\alpha_{xy}(t)$. The noise power $S(f)$ of a signal $x(t)$ is defined as the Fourier transform of the temporal correlation function,
\begin{equation}
S(t-t')\equiv\frac12\left\langle \Delta x(t) \Delta x(t') + \Delta x(t') \Delta x(t)\right\rangle,\label{stt}
\end{equation}
where $\Delta x(t)\equiv x(t) - \left\langle x\right\rangle$.

Depending on the dynamical regimes described above, $S(f)$ shows different qualitative behaviors. We note here that in the $\kappa \rightarrow \infty$ limit, the shear modulus of the vortex matter vanishes and thus only the vortex liquid phase survives.~\cite{glazman_thermal_1991} To give a full picture, we provide a short description of the noise behavior in the finite-$\kappa$ regime also, but a more detailed account will be published elsewhere. In the pinned phase, the Nernst signal is suppressed and the noise power is nearly independent of $f$, thus displaying a white-noise spectrum. In the sliding-lattice phase, the noise power shows a characteristic narrow-band behavior and most of the spectral density is found near a particular frequency. The characteristic frequency can be estimated as $f_{\text{w}} \sim v/ \ell$, where $v$ is the speed of vortices and $\ell$ is the distance between vortices in the direction of motion.  This type of noise is known as washboard noise and has been observed experimentally in the conduction noise spectra of driven BSCCO samples.\cite{togawa_direct_2000,*maeda_experimental_2002}  In the vortex-liquid regime, noise power again shows a wide-band behavior. In the infinite-$\kappa$ limit, we found it to be fitted well by a power law $1/f^\beta$.
\begin{figure}[!htb]
\includegraphics[width=3.3in]{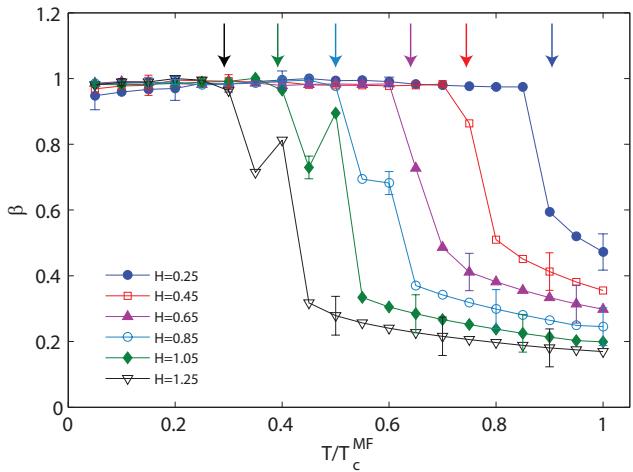} 
\caption{(Color online)  Exponent of the noise power, $\beta$, as a function of $T$ for different values of $H$.  The error bars (shown for selected data points) have been obtained from a least-squares fitting procedure on the log \emph{S} vs log $f$ plots.  The vertical arrows mark the onset temperatures ($T_{\mathrm{NN}}$) for the negative Nernst signal obtained using linear interpolation in Fig. 3(a).}
\label{Noise} 
\end{figure}
Figure \ref{Noise} shows the plot of $\beta$ for different magnetic fields as a function of temperature. We observe that in the low-$T$ vortex-liquid regime, the exponent $\beta \approx 1$ and is nearly independent of the applied magnetic field. As the system enters into the negative-Nernst-signal regime, $\beta$ drastically drops to lower values. This sharp feature provides an excellent means for experiments to causally correlate the onset temperature for negative Nernst signals, $T_{\text{NN}}$, to vortex dynamics. As the temperature approaches the mean-field transition temperature, the exponent is further reduced, but remains finite. In contrast to the low-$T$ vortex-liquid regime, in this case, $\beta$ strongly depends on the applied magnetic field. We remark that we do not see any observable signatures of the vortex-proliferation transition in the behavior of $\beta$. Even though $1/f$ behavior of noise is ubiquitous in transport measurements of many physical systems,\cite{dutta_low-frequency_1981,*weissman_1/f_1988} to the best of our knowledge, it has not been measured for the Nernst signal. In YBCO samples, a similar power-law behavior has been observed in flux noise spectra in a nondriven system~\cite{vestergren_simulation_2004,*festin_vortex_2004} and in longitudinal-voltage noise spectra in a current-driven system.\cite{giraldo_non_2009} Although the above measurements have been done for regimes and observables other than those presented here, they illustrate the feasibility of the noise-measurement scenario we propose here.

\section{Closing remarks}
In this article, a computational method has been introduced to simulate the Nernst effect in high-$T_c$ superconductors using the TDGL theoretical framework.  We show that the thermoelectric coefficient $\alpha_{xy}$ (i) is large and monotonically decreasing for $T<T_{\mathrm{VP}}$, (ii) displays a profile that is qualitatively consistent with experiments at intermediate $T$'s, and (iii) reverses signs at $T>T_{\mathrm{NN}}>T_{\mathrm{VP}}$.  Furthermore, we propose a noise-based observable to distinguish between different dynamical phases of the vortex system.  The noise power of the Nernst signal reveals distinctive qualitative and quantitative properties for different regimes, thus demonstrating its great potential as an experimental probe of vortex dynamics.

\begin{acknowledgments}
We gratefully acknowledge discussions with H.~Steinberg on $1/f$ noise, and with T.~Giamarchi, S.~Mukerjee, D.~Podolsky and S.~Tewari on vortices and the Nernst effect. We also acknowledge the hospitality of the Aspen Center for Physics and the Kavli Institute for Theoretical Physics where some of these discussions took place (Grant No.~NSF PHY06-02228 and No.~PHY05-51164, respectively).  This work was partially supported by the W.~M.~Keck and Robert A.~Welch (C-1681) Foundations.
\end{acknowledgments}

\appendix
\section*{Appendix}
\begin{figure}[!htb]
\includegraphics[width=3.3in]{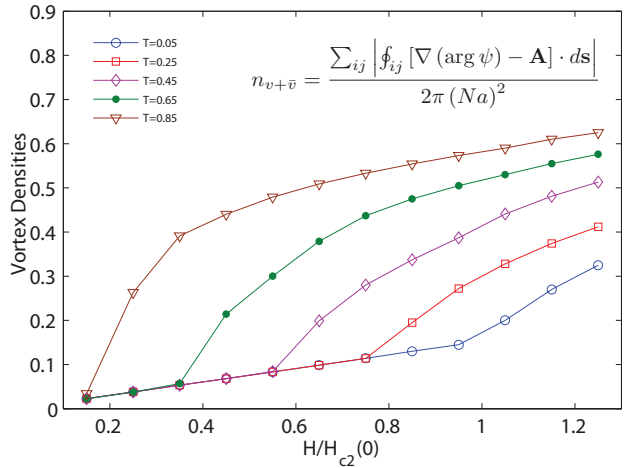} 
\caption{(Color online) Magnetic field dependence of $n_{v+\bar{v}}$, the total density of vortices plus antivortices, showing the onset of the vortex-proliferation regime.}
\label{VorticityVsH} 
\end{figure}

For ease of comparison with the existing literature, we alternatively present the data in Figs. \ref{Alphaxy}(a) and \ref{Vortices}(b) as functions of magnetic fields for several temperatures [contrary to functions of temperatures for several fields as in Figs. \ref{Alphaxy}(a) and \ref{Vortices}(b)].  For any given curve in Fig.~\ref{VorticityVsH}, we see a linear increase of vortex densities until reaching a certain field value, at which vortices proliferate due to the onset of vortex-antivortex pair generation.  The proliferation starts at higher fields for lower temperatures.

\begin{figure}[!htb]
\includegraphics[width=3.3in]{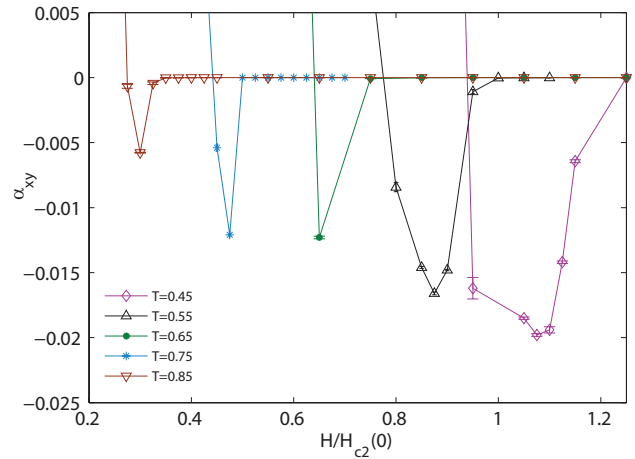} 
\caption{(Color online)  Magnetic field dependence of $\alpha_{xy}$; the error bars (comparable to symbol sizes) highlight the statistical accuracy of the negative signals.  The error bar for each data point has been obtained from eight independent measurements of $\alpha_{xy}$, each measurement consisting of time averaging over 0.1 million time steps.}
\label{alpha_xy_vs_H_negative} 
\end{figure}

Figure \ref{alpha_xy_vs_H_negative} shows the plot of $\alpha_{xy}$ in the region where the negative values are conspicuous.  As $H$ is increased beyond $H_{c2}(T)$ for a given temperature, $\alpha_{xy}$ reverses sign and becomes negative before diminishing towards zero from the negative side.  The sign reversal of $\alpha_{xy}$ occurs at higher magnetic fields and the magnetic-field ranges for negative $\alpha_{xy}$ broaden for lower temperatures.

\bibliographystyle{apsrev4-1}
\bibliography{bibliox}
\end{document}